\documentclass[11pt,a4]{article}
\usepackage{times,fancyhdr}
\usepackage[dvips]{graphicx}
\usepackage{array}
\usepackage{amssymb}
\usepackage[small,bf]{caption}
\usepackage{color}
\usepackage{fancyhdr}
\usepackage{natbib}
\usepackage{setspace}
\usepackage{framed}
\usepackage{xcolor}
\usepackage{verbatim,framed}
\usepackage[yyyymmdd,hhmmss]{datetime}
\fancyheadoffset[R]{0pt}
\fancyfootoffset[R]{0pt}
\renewcommand{\headrulewidth}{0.4pt}
\renewcommand{\footrulewidth}{0.4pt}
\definecolor{orange}{rgb}{1.0,0.5,0.0}
\definecolor{aqgr}  {rgb}{0.0,1.0,0.6} % aqua green
\definecolor{viol}  {rgb}{0.8,0.6,0.8}
\definecolor{figdr} {rgb}{1.0,1.0,1.0} % 0.8,0.6,0.7
\definecolor{coldr} {rgb}{1.0,0.0,1.0} % fuchsia
\definecolor{colhd} {rgb}{1.0,0.8,0.0} % gold

\newcolumntype{C}[1]{>{\centering\let\newline\\\arraybackslash\hspace{0pt}}m{#1}}
\title{\bfseries{\textsc{a theoretical model of soma-to-germline \\
transmission of transposable elements \\
to build new gene regulatory sequences}}}

\author{Alessandro Fontana (AF), PhD \\
email address: fontalex00@gmail.com}

\begin{document}
\maketitle
   
\clubpenalty=20000
\widowpenalty=20000

\begin{abstract}
Transposable elements are DNA sequences that can move around to different positions in the genome. During this process, they can cause mutations, and lead to an increase in genome size. Despite representing a large genomic fraction, transposable elements  have no clear biological function. This work builds upon a previous model, to propose a new concept of natural selection which includes a non-Darwinian component. Transposable elements are hypothesised to be the vector of a flow of genetic information from soma to germline, that shapes gene regulatory regions across the genome. The paper introduces the concept, presents and discusses the body of evidence in support of this hypothesis, and suggests an experiment to test it. 
\end{abstract}

%\pagebreak[4]

\section{Introduction}

%\colorbox{colhd}{transposable elements} \\
\textit{Transposable elements (TE)} \cite{Mcclintock50}, are DNA sequences that can move around to different positions in the genome. During this process, they can cause mutations, chromosomal rearrangements and lead to an increase in genome size. Despite representing a large genomic fraction (30-40\% in mammals), TEs have no clear biological function. TEs are active during development, and may bring diversity in somatic cells having the same genome \cite{Collier07}. A TE activity was also observed in germline cells, with possible evolutionary implications. ``Waves'' of TE diffusion in a genome appear temporally associated with major evolutionary changes to the species \cite{Oliver10}, suggesting a causal link between the two events.

%\colorbox{colhd}{paper structure} \\
This work suggests a hypothesis on the role of TE that builds on previous proposals \cite{Fontana10c, Fontana12b}, and based on a model of embryonic development called \textit{Epigenetic Tracking} \cite{Fontana08}. While the study of life sciences often relies on a bottom-up method, that tries to infer general rules starting from genes and proteins, our work is informed by a top-down approach, that proceeds from high-level abstractions towards a proposal for biological, low-level molecular processes. As a consequence our model may contain ingredients not necessarily adherent to current knowledge, but which can become a suggestion for biologists to look into new, previously unexplored directions. This paper is organised as follows: this (first) section is the introduction; section 2 describes the model of development; section 3 is dedicated to the evo-devo process and the role of TEs; section 4 contains the discussion; section 5 draws the conclusions and outlines future directions.

\section{Development}

\begin{figure}[t] \begin{center} \hspace*{-0.0cm}
{\fboxrule=0.0mm\fboxsep=0mm\fbox{\includegraphics[width=12.00cm]{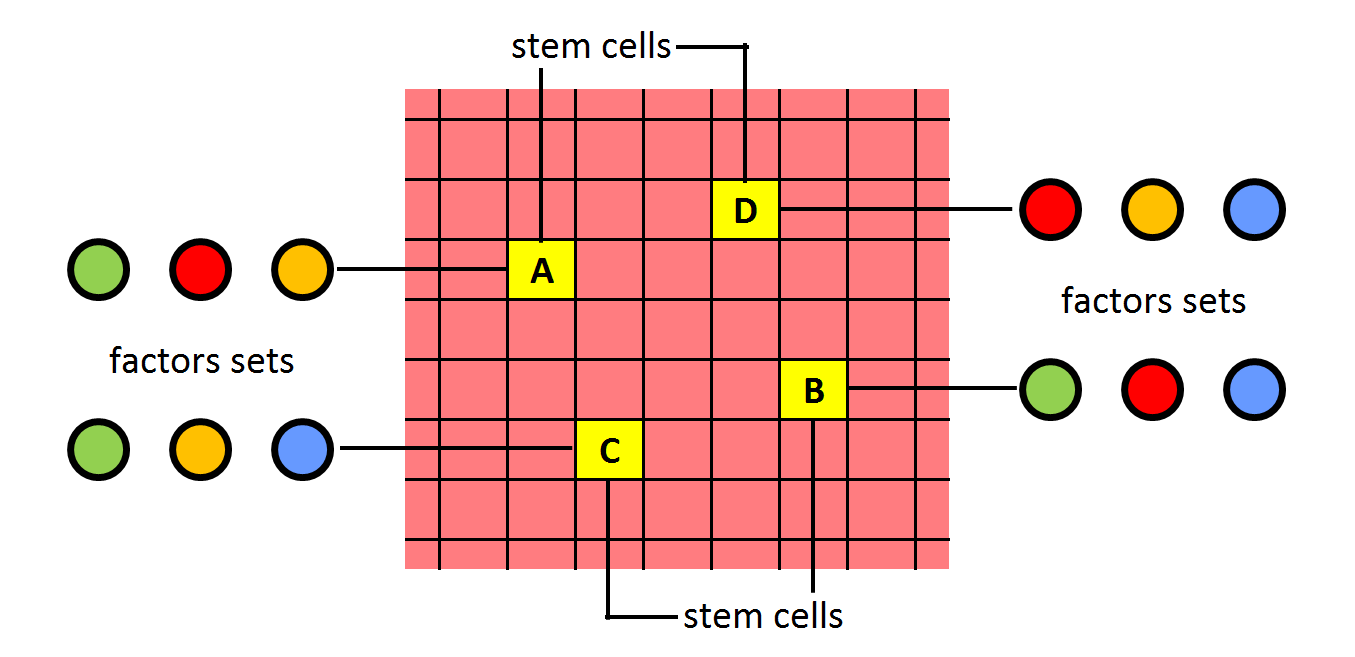}}}
\caption{Stem cells and master transcription factors. Stem cells (in yellow) are endowed with a subset of factors called factor set (in this example composed of three factors). These factors are responsible for the activation of batteries of genes downstream which determine cellular behaviour. Different factor sets are shown with different letters.}
\label{reperta}
\end{center} \end{figure}

%\colorbox{colhd}{cells and factor set} \\
In our model development starts from a single cell and unfolds in a predefined number of developmental stages. Organisms are composed of two categories of cells: \textit{normal cells} and \textit{stem cells}. \textit{Master transcription factors} (for brevity \textit{factors} from now on) are responsible for the activation of batteries of genes downstream which in the end determine cellular behaviour. Examples of factors active during early embryogenesis are represented by Hox proteins. Each stem cell has an associated \textit{factor set}, composed of a specific subset of factors (Fig.~\ref{reperta}).

\begin{figure}[t] \begin{center} \hspace*{-0.75cm}
{\fboxrule=0.0mm\fboxsep=0mm\fbox{\includegraphics[width=18.00cm]{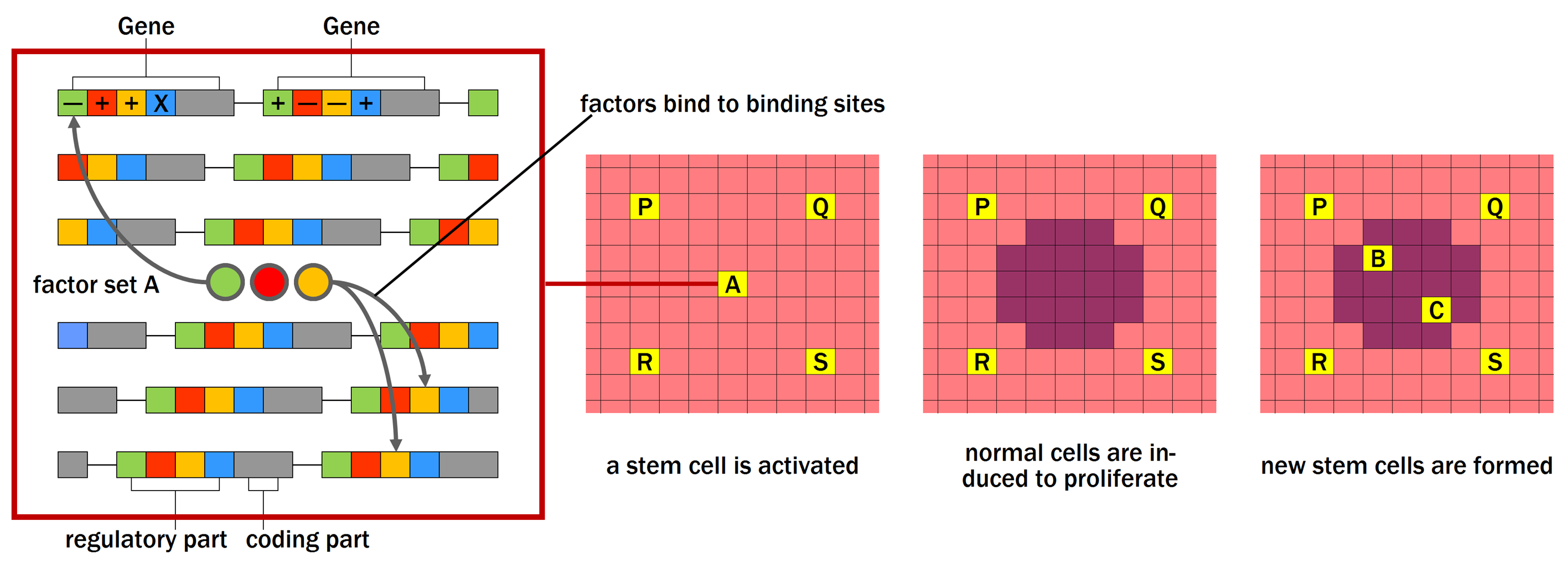}}}
\caption{Developmental events. The genome is structured as an array of genes, each composed of a coding part (shown in grey) and a regulatory part, represented as a list of as many binding sites as the number of factors in the organism's repertoire. Each binding site is shown in the same colour as the corresponding factor, and is associated to a symbol (+,-,x) which defines its type of contribution to gene transcription. The cascade of genes triggered by the factor set can cause the stem cell to get activated. Upon activation, the stem cell (with factor set A, in the left grid panel), can induce normal cells to proliferate (middle grid panel) or to differentiate (not shown). Subsequently, some normal cells revert to stem cells, and obtain new factor sets (B, C in right grid panel).}
\label{events}
\end{center} \end{figure}

%\colorbox{colhd}{genome} \\
Cells have a genome, structured as an array of genes (Fig.~\ref{events}). Each gene is composed of a coding part and a regulatory part, represented as a list of as many binding sites as the number of factors in the organism's repertoire. Each binding site is characterised by the type of contribution it gives to gene activation: factor presence required for activation (+), factor absence required for activation (-), factor presence irrelevant for activation (x). When a gene is activated, it produces a protein that can become a transcription factor for other genes. 

%\colorbox{colhd}{events} \\
Stem cells direct development. When the proteome generated by the cascade of genes triggered by the factor set causes the stem cell to get activated, a specific \textit{developmental event} is orchestrated (Fig.~\ref{events}): normal cells present in the volume nearby are induced to proliferate or to differentiate. If no normal cells are present in the volume, they can be generated by the stem cell through several rounds of asymmetric cell divisions (i.e. the stem cell produces another identical stem cell and a normal cell). 

%\colorbox{colhd}{stem formation} \\
Subsequently, some normal cells revert to stem cells, and obtain new factor sets. Some of these newly formed stem cells can become the centre of other developmental events at a later stage, and development can move ahead. Hence, in our model the normal-stem transition is a two-way street. This idea is consistent with recent models of stem cells, such as the ``complex system model'' \cite{Cruz12}, or the ``stemness phenotype model'' \cite{Laks10}. These models share the idea that stemness is a dynamic property: the stem-non stem conversion would occur in both directions, triggered by genetic and epigenetic factors, and influenced by the cellular microenvironment.

\begin{figure}[t] \begin{center} \hspace*{-0.0cm}
{\fboxrule=0.0mm\fboxsep=0mm\fbox{\includegraphics[width=15.50cm]{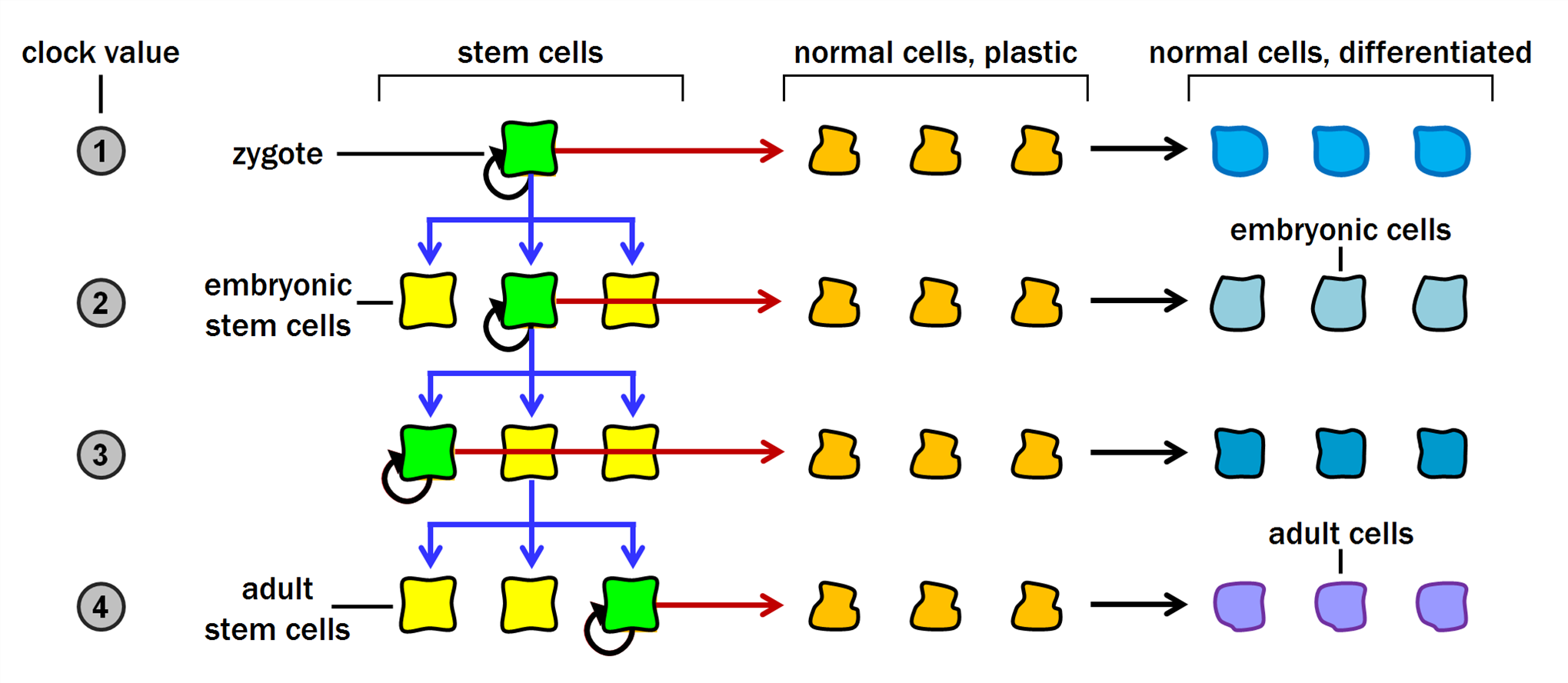}}}
\caption{Set of stem cells created during development. Stem cells marked in green become active during development. As a result, they induce the prolifeartion and differentiation of normal cells. In parallel, other stem cells are generated.}
\label{biocells}
\end{center} \end{figure}

%\colorbox{colhd}{tree of stem cells} \\
The set of stem cells generated during development has a hierarchical, or tree, structure (Fig.~\ref{biocells}). This feature appears consistent with the information we have on the set of biological stem cells involved in the generation of particular organs or systems, such as the hematopoietic system \cite{Reya01}. Our model provides also a means to bridge the conceptual gap between embryonic and adult stem cells: embryonic stem cells correspond to elements in the tree near the root, while adult stem cells correspond to the leaves of the tree. It also explains how adult stem cells are generated, with the mechanism of conversion from normal cells. 

%\pagebreak[4]
%\begin{figure}[p] \begin{center}
%{\fboxrule=0.0mm\fboxsep=0mm\fbox{\includegraphics[height=16.00cm]{xxblank}}}
%\end{center} \end{figure}

%\clearpage
%\pagebreak[4]

\section{Evolution and Germline Penetration}
\label{sec:evo-devo}

\begin{figure}[t] \begin{center} \hspace*{-0.50cm}
{\fboxrule=0.0mm\fboxsep=0mm\fbox{\includegraphics[width=17.50cm]{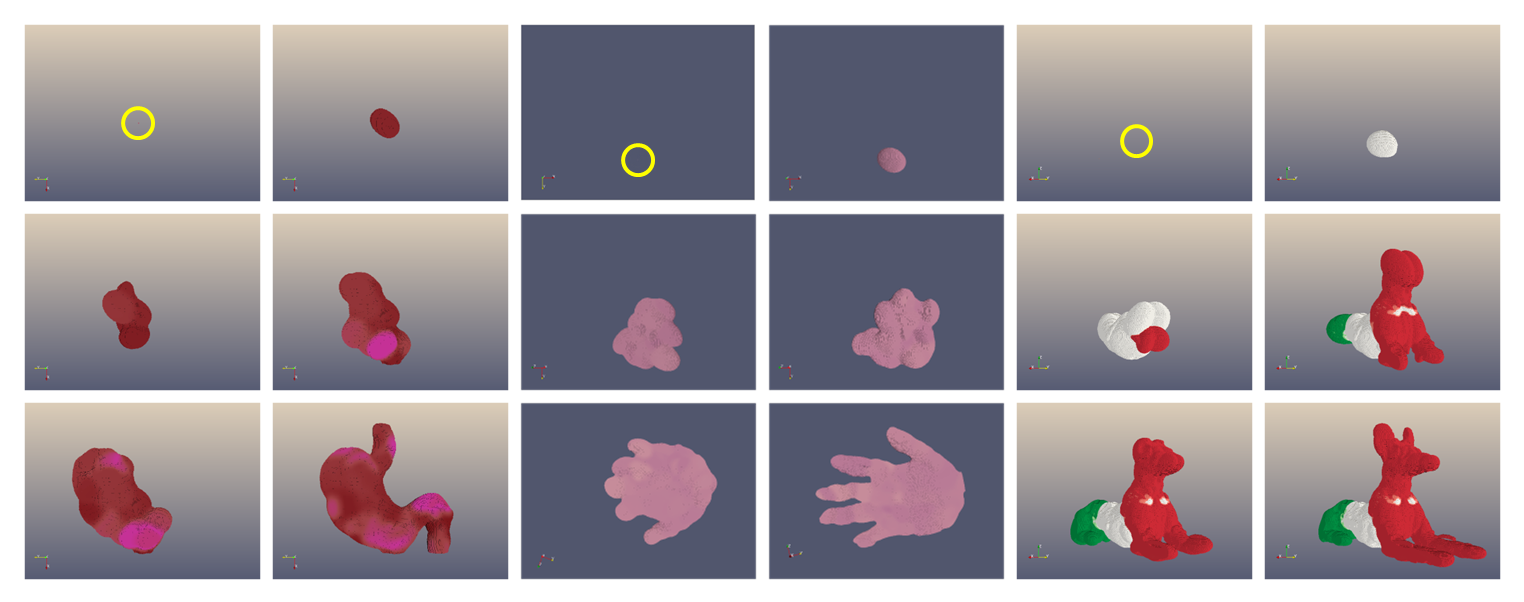}}}
\caption{Examples of structures generated using our model. The panels show some stages of development from the single cell stage (yellow circles) for three structures. The structures, composed of millions of cells, represent a stomach, a hand, and a dog patterned with the colours of the Italian flag. This last one is a more complex version of the classical Wolpert's French frag model.}
\label{devsimul}
\end{center} \end{figure}

%\colorbox{colhd}{evo-devo method and results} \\
The model of development described can be coupled with a \textit{genetic algorithm}, able to simulate Darwinian evolution. The genetic algorithm evolves a population of individuals (each encoded in an artificial genome) for a number of generations. At each generation, all individuals develop independently from the zygote stage to the final phenotype, whose proximity to a predefined target is employed as a fitness measure. This operation is repeated for all individuals, so that eventually each individual is assigned a fitness value. Based on this value the genomes of the individuals are selected and randomly mutated, to produce a new population. This cycle is repeated until a satisfactory level of fitness is reached. The coupling of the model of development and the genetic algorithm gives origin to an evo-devo process, which was proved able through computer simulations to ``devo-evolve'' 3-dimensional structures of unprecedented complexity (see examples in Fig.~\ref{devsimul}). 

\begin{figure}[t] \begin{center} \hspace*{-0.50cm}
{\fboxrule=0.0mm\fboxsep=0mm\fbox{\includegraphics[width=17.50cm]{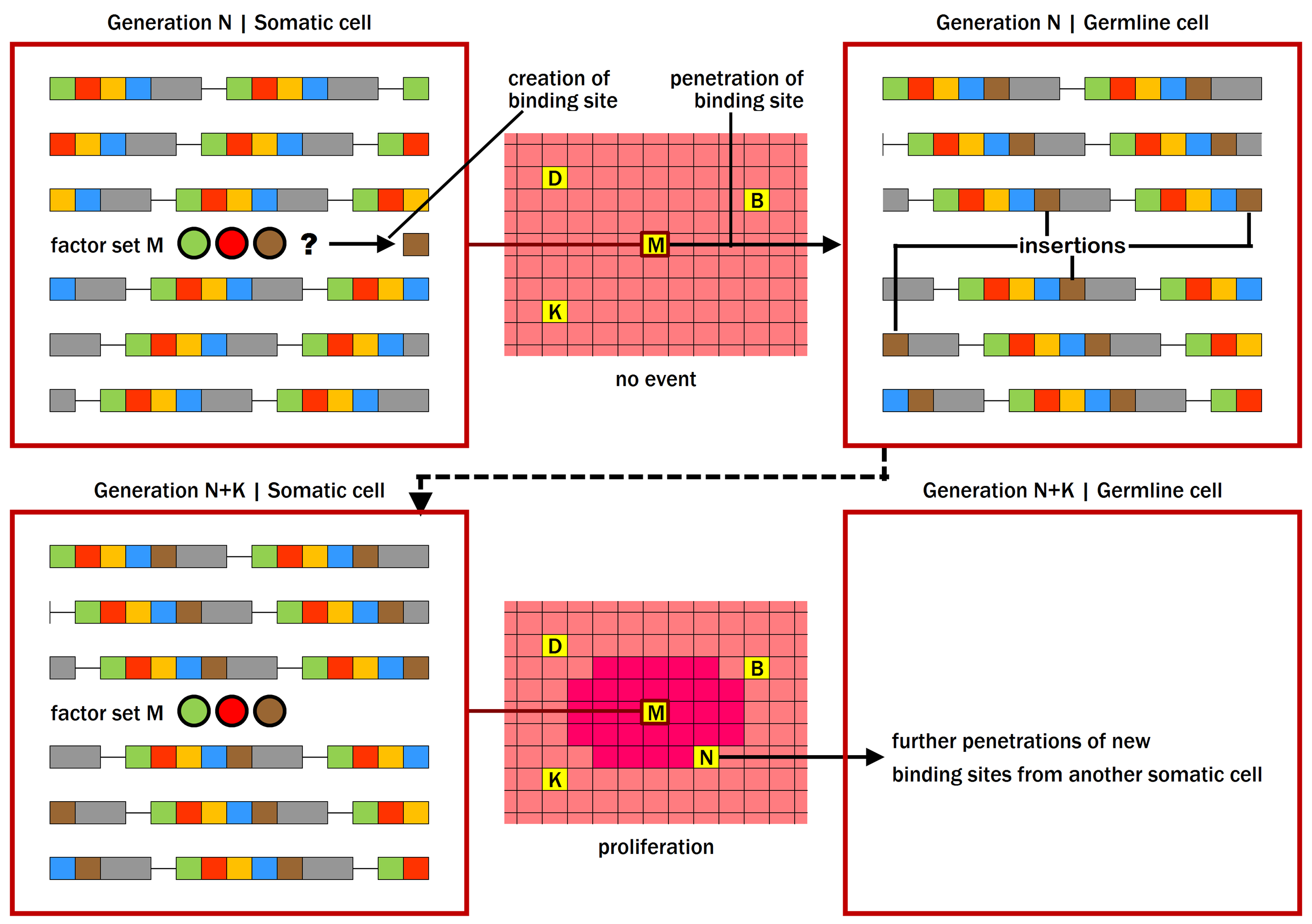}}}
\caption{Germline penetration of new binding sites in our model. In the upper part, in a stem cell with factor set M, the genome is not able to ``respond'' to factor set M with a corresponding event, because the genome lacks binding sites for the brown factor. A new binding site for the brown factor is created in the stem cell. The new binding site leaves the stem cell and reaches the germline, where it is incorporated in the genome in the regulatory part of each gene, next to other sites. In a subsequent generation, when a cell has factor set M, its genome (inherited from the penetrated germline) can respond with an event.}
\label{noevent}
\end{center} \end{figure}

%\colorbox{colhd}{GP, problem and description} \\
In our model most stem cells produced during development do not orchestrate any events (are inactive), because in the genome there is no gene whose regulatory part contains the binding site for one or more factors (upper left part of Fig.~\ref{noevent}). The probability that a suitable binding site emerges in the genome simply through mutations and recombinations is very low. A countermeasure consists in ``suggesting'' to the genetic algorithm to put in the genome binding sites which are guaranteed to match. This idea is implemented in a procedure called \textit{germline penetration}, which builds binding sites able to match factors generated during development. These binding sites are then inserted next to genes contained in a special copy of the genome called ``germline'' genome (upper right part of Fig.~\ref{noevent}) which, after reproduction, is destined to become the (``somatic'') genome of the individuals of the next generation. 

%\colorbox{colhd}{GP, example} \\
Once evolution is provided with ``good'' binding sites, guaranteed to match an existing factor, it has to optimise the contribution fields (+,-,x) and possibly also the coding part of the associated gene, a process that can take several generations. When this optimisation is completed, some genes can be activated in the spatio-temporal context of stem M and a developmental event is produced (lower part of Fig.~\ref{noevent}). New stem cells are generated as a result of the new event, including new factors: binding sites able to match these factors may again be transferred to the germline genome, and the whole cycle repeats itself.
 
%\colorbox{colhd}{GP is essential, what about nature?} \\
Germline penetration is essential in the model for the evolution of complex shapes: in our \emph{in silico} experiments, if germline penetration is disabled, the evolutionary process practically grinds to a halt. The central role played by germline penetration in our model lead us to hypothesise the existence of a similar process also in biological organisms. For the implementation of a biological germline penetration, we need genetic elements able to build new binding sites in somatic cells, and susceptible of being transferred to germline cells: TEs seem to possess all characteristics to play this important biological role. 

%\colorbox{colhd}{temporary sequences generated} \\
Based on these considerations, we can imagine a possible implementation of a biological germline penetration. In a stem cell unable to produce a developmental event (because some factors present in the cell cannot find corresponding binding sites), TEs are recruited, to build temporary binding sites, able to match the existing factors. Then, the TEs leave the cell and, through the bloodstream, make their way to the germline. Here they insert themselves into the genome, thus allowing the delivery of a successful innovation to the next generation.

%\pagebreak[4]
%\begin{figure}[p] \begin{center}
%{\fboxrule=0.0mm\fboxsep=0mm\fbox{\includegraphics[height=16.00cm]{xxblank}}}
%\end{center} \end{figure}

%\clearpage
%\pagebreak[4]

\section{Discussion}

%\colorbox{colhd}{mold hypothesis/1} \\
In essence, our hypothesis is that a given factor acts as a ``mold'' for the creation of its own binding site which, carried on a TE, is then spread in multiple copies across the genome. We argue that, given the evidence, this is the only possible scenario. The evidence is that, for a given factor, the genome hosts a large number of TE-derived binding sites. If we exclude the ``mold'' hypothesis, how could these binding sites be created? 

%\colorbox{colhd}{mold hypothesis/2} \\
The binding site must first be created in the germline in a single copy, and then it must spread across the genome: this is the only possible explanation for the fact that the binding sites are all similar to each other. Since a binding site is composed on average of 6-10 base pairs, there are a total of $4^{10}$ = 1 million possible binding sites. The testing of all such combinations through Darwinian evolution would require a huge number of generations, and this for a single factor. This is not compatible with the observed speed of evolution. 

%\colorbox{colhd}{mold hypothesis/3} \\
If we accept the ``mold'' hypothesis, there must be a place in which the factor and the TE are present at the same time, so that the factor can serve as a mold to create, by mechanisms to be elucidated, its own binding site on the TE. And this can only happen in a somatic cell. As a matter of fact, while TEs are present in both somatic and germline cells, the factor is produced only in specific somatic cells. The TE-binding site must then be transferred to the germline, to be replicated in multiple copies across the genome and produce trans-generational effects.  

%\colorbox{colhd}{waves} \\
The mechanism described has interesting evolutionary implications. In our model, whenever a stem cell proliferates, a wave of new stem cells, with new factor sets, is created in the body of the (new) species. The action of germline penetration translates this wave of new factor sets in somatic cells into a corresponding wave of new binding sites spreading in the germline genome and subsequently in the somatic genome of future generations. Such events during evolution coincide with moments in which major changes occur to the evolving species, causing new body parts or features to appear. In biological terms, this means that the spreading in the genome of waves of new TE sets in the course of evolution correspond to moments in which new branches (new species) are generated in the ``tree of life''. Such predictions, made with our model on purely theoretical grounds, appear to be confirmed by experimental evidence \cite{Oliver10}. 

%\colorbox{colhd}{Waves and timing} \\
This seems to hint that the TE colonisation of the genome is the driving force behind the change in the lineage. Our interpretation of this phenomenon is different. In fact, according to our model, the spread of new TE families in the genome is an event which comes immediately (in evolutionary terms) {\em after} the change, not before. The confirmation of this prediction may be obtained through techniques to estimate the age of DNA sequences more sensitive than those currently employed. Another prediction of this model is that major waves of TE colonisation during evolution coincide with the emergence of key genes-transcription factors in the genome, such as genes of the Hox cluster.

%\colorbox{colhd}{TE and binding sites} \\
The notion that TEs may provide a rich source of genetic material from which gene binding sites can evolve is well established \cite{Jacques13, Rebollo12}, and the hypothesis that TE-shaped binding sites were evolved to accomodate master transcription factors (such as p53) has already been entertained \cite{Feschotte08}. Our contribution is to fit this observation into a new paradigm of evolution and development, which considers an inverse flow of genetic information as an essential component. 

%\colorbox{colhd}{Plos and SMGT} \\
TEs are involved in a process by which spermatozoa take up exogenous DNA molecules and deliver them to oocytes at fertilisation \cite{Smith05}. The reverse-transcribed molecules are incorporated into the germline genome and become part of the hereditary material which is transferred from one generation to the next. Experimental evidence suggests that this kind of gene transfer is a retrotransposon-mediated phenomenon. The ability of spermatozoa to take up DNA molecules and incorporate them into their genome could represent the mechanism used by nature to implement the final stage of a biological germline penetration. A recent study showed that somatic cells can release genetic material that can later be found in the DNA of the offspring \cite{Cossetti14}. The hypothesised migration of TEs from somatic to germ cells seems therefore to be coherent with current experimental evidence.

%\colorbox{colhd}{rise, maize, barley} \\
Despite being rice, maize and barley very similar species, they differ widely in terms of genome size, ranging from 350 MB (rice) through 2500 MB (maize) to 5200 (barley) \cite{Feschotte02}. While the number and structure of coding genes appear to be very similar among the three species, the mean distance between adjacent genes is very different. The increase of this distance, which is proportional to the genome size in each species, can be traced back to massive TE insertions in inter-genic regulatory regions. This piece of evidence appears consistent with the model proposed.

%\colorbox{colhd}{view of evolution} \\
In the perspective suggested, coding genes remain essentially unchanged during evolution, both in number and in structure. An interpretation of this phenomenon is that the set of existing coding genes are able to generate a number of protein combinations sufficient to build any kind of biological structures. Once evolution is provided with a rich repertoire of building-blocks, it may find more useful to discover new ways in which these components can be assembled, in many different spatio-temporal contexts (of course this does not preclude the emergence of new coding genes).   

%\colorbox{colhd}{why jumping genes} \\
According to our hypothesis, evolution continuously adds binding sites to protein-coding genes. At any moment during evolution, the space between adjacent genes is dedicated to hosting binding sites for existing factors: only factors that are allowed to bind in these regions can influence gene activity. Therefore, the new binding sites cannot be added at the ``end'' of the genome (whatever it may mean), but must be put next to the gene they are intended to control. The need to use for this task mobile elements, able to insert into random genomic locations, appears evident. 

%\colorbox{colhd}{no precise insertions are required /1} \\
Compared to our previous proposals \cite{Fontana10c, Fontana12b}, this model version is much more adherent to the reality of biological cells: genes are activated when trascription factors find their corresponding binding sites, as it happens in nature. This innovation has an important implication on the dynamics of TE diffusion. In our previous models, the regulatory parts of genes were monolithic blocks. As a result, germline penetration had to move a complete regulatory region from a position of the somatic genome to the very same position of the germline genome. This implied the ability of TEs to insert themselves into precise genomic locations, a feature which is not supported by experimental evidence. 

%\colorbox{colhd}{no precise insertions are required /2} \\
In the current model, thanks to the subdivision of the regulatory part into individual binding sites, the sequence corresponding to a single binding site is only required to spread copies across all (or most) gene regulatory parts. As a result, the TE insertions need only occur in ``hot spots'' regions, located in the vicinity of genes, a fact which is corroborated by a large body of evidence \cite{Lowe07}. This makes the foreseen TE dynamics much more biologically plausible.

%\colorbox{colhd}{experiment proposal} \\
We wish to conclude our discussion by proposing an experiment to test the theory. The experiment consists in inserting into an individual of species S a transgene X, which: i) is expressed at a late stage of development and ii) whose product x does not have binding sites in the genome. If germline penetration works as expected, we should be able to find newly formed binding sites for x in individuals of subsequent generations, assembled with one or more TEs.

%\pagebreak[4]
%\begin{figure}[p] \begin{center}
%{\fboxrule=0.0mm\fboxsep=0mm\fbox{\includegraphics[height=16.00cm]{xxblank}}}
%\end{center} \end{figure}

%\clearpage
%\pagebreak[4]

\section{Conclusions}
  
%\colorbox{colhd}{conclusions} \\
In this work we have proposed an extension of our model of embryonic development that is much closer to the reality of biological cells. This allowed us to propose a model of germline penetration of binding sites that is more biologically plausible in terms of TE dynamics, and accounts for current experimental evidence. Furthermore, we have proposed an experiment to test the correctness of this idea. Future work will be aimed at further reducing the gap between the model and biological reality.

\singlespacing
\bibliographystyle{plain}
\bibliography{ldanxtrans}
 
\end{document}